\documentclass{PoS}
\newcommand{\bp}{\mbox{$\vec{p}_T$}}
\newcommand{\bq}{\mbox{$\vec{q}_T$}}
\newcommand{\qperp}{\mbox{$\vec{q}_\perp$}}
\newcommand{\eperp}{\mbox{$\vec{e}_\perp$}}
\newcommand{\aem}{{\alpha_{\mathrm{em}}}}

\title{Production of $W^+ W^-$ and $t \bar t$ pairs 
via photon-photon processes 
in proton-proton collisions}

\ShortTitle{Production of $W^+ W^-$ and $t \bar t$ pairs}

\author{\speaker{Antoni Szczurek}\thanks{A footnote may follow.}\\
        Institute of Nuclear Physics PAN, Krak\'ow\\
        E-mail: \email{antoni.szczurek@ifj.edu.pl}}

\author{Marta Luszczak\\
        Rzesz\'ow University, Rzesz\'ow\\
        E-mail: \email{luszczak@ur.edu.pl}}

\abstract{We review our recent results for production of
$W^+ W^-$ and $t {\bar t}$ pairs via photon-photon fusion.
A theoretical approach is presented in short.
We include transverse momenta of photons when calculating fluxes of photons.
Then we discuss our results for cross section (total and differential) 
for $W^+ W^-$ production. 
Results for different parametrizations of proton structure
functions are used to calculate inelastic fluxes of photons.
A discussion on rapidity gap survival probability
due to remnant fragmentation is presented. 
A similar discussion is presented for $t {\bar t}$ production.}

\FullConference{XXVII International Workshop on Deep-Inelastic Scattering and Related Subjects - DIS2019\\
		8-12 April, 2019\\
		Torino, Italy}

\begin{document}

\section{Introduction}

It was realized rather recently that the electroweak corrections 
are important for precise calculations of cross sections in different 
processes. The $p p \to W^+ W^-$ process is a good example
(see e.g. \cite{LSR2015}). Then $\gamma \gamma \to W^+ W^-$ is the most
important subprocess. This subprocess is important also in the context
of searches beyond Standard Model \cite{Chapon:2009hh,Pierzchala:2008xc}.
By imposing special conditions on the final state this contribution can 
be observed experimentally \cite{Khachatryan:2016mud,Aaboud:2016dkv}.

In \cite{daSilveira:2014jla,Luszczak:2015aoa} we developed 
a formalism for calculating
$p p \to l^+ l^-$ processes proceeding via photon-photon fusion.
In \cite{LSS2018} we used the same technique to calculate
cross section for $p p \to W^+ W^-$ reaction proceeding via
photon-photon fusion. In order to make reference to real
``measurements'' of the photon-photon contribution one has to include
in addition the gap survival probability caused by extra emissions.
In \cite{LFSS2019a} we concentrated on the effect related 
to remnant fragmentation and its destroying of the rapidity gap.

In \cite{LFSS2019b} we calculated cross section for the photon-photon
contribution for the $p p \to t \bar t$ reaction including also effects 
of gap survival probability.

Here we briefly review our results obtained in 
\cite{LSS2018,LFSS2019a,LFSS2019b}.

\section{Our approach}

In our analyses we included different categories of processes shown in
Fig.\ref{fig:diagrams}.

\begin{figure}
\begin{center}
\includegraphics[width=5cm,height=3.6cm]{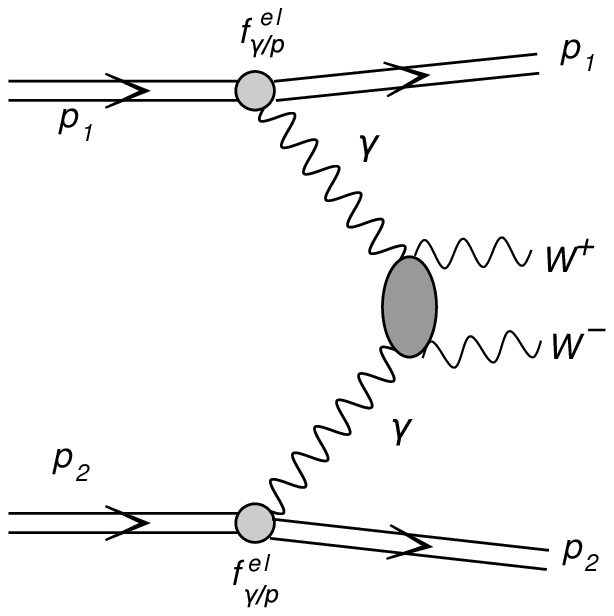}
\includegraphics[width=5cm,height=3.6cm]{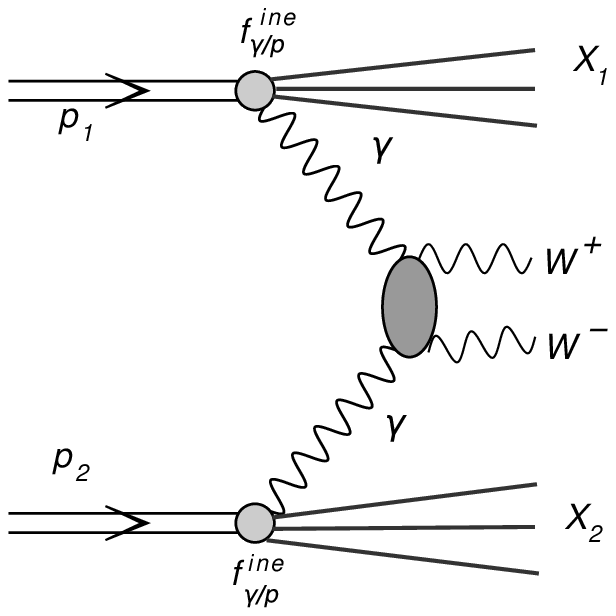} \\
\includegraphics[width=5cm,height=3.6cm]{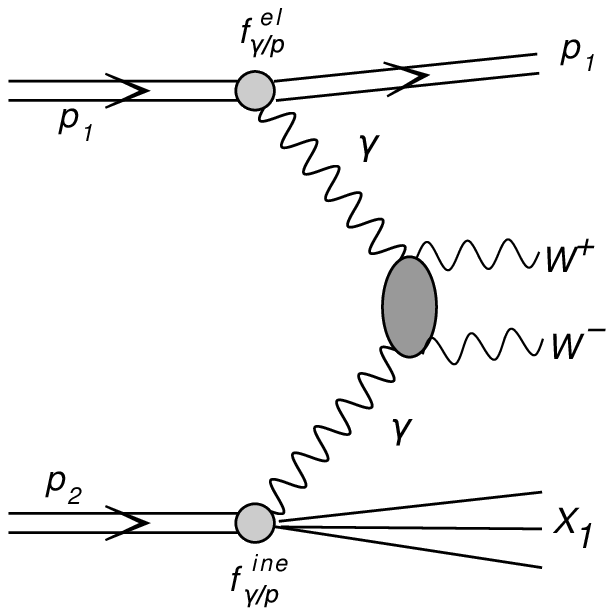}
\includegraphics[width=5cm,height=3.6cm]{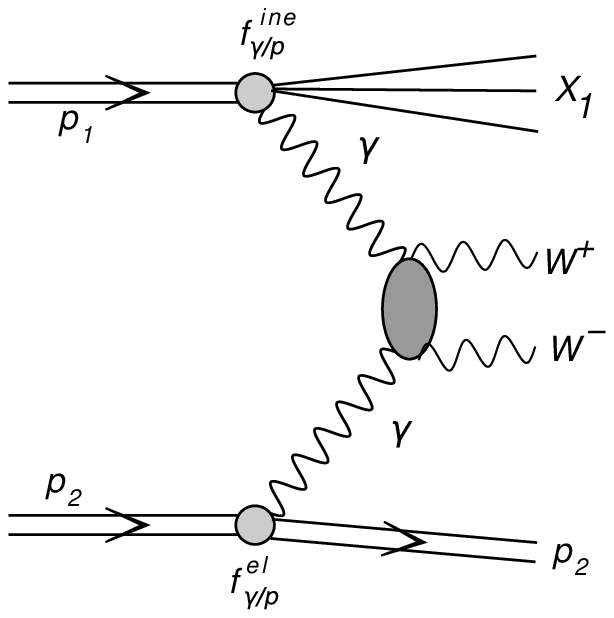}
\caption{Diagrams representing different categories of photon-photon 
induced mechanisms for production of $W^+ W^-$ pairs.
}
\label{fig:diagrams}
\end{center}
\end{figure}

In contrast to other authors, in our approach we include transverse
momenta of (virtual) photons.
Then the differential cross section for $W^+$ and $W^-$ production can
be written as:

\begin{eqnarray}
{d \sigma^{(i,j)} \over dy_1 dy_2 d^2\bp_1 d^2\bp_2} &&=  \int  {d^2 \bq_1 \over \pi \bq_1^2} {d^2 \bq_2 \over \pi \bq_2^2}  
{\cal{F}}^{(i)}_{\gamma^*/A}(x_1,\bq_1) \, 
{\cal{F}}^{(j)}_{\gamma^*/B}(x_2,\bq_2) 
{d \sigma^*(p_1,p_2;\bq_1,\bq_2) \over dy_1 dy_2 d^2\bp_1 d^2\bp_2} \, ,
\label{eq:kt-fact}
\end{eqnarray}
where $i,j$ = elastic, inelastic and the longitudinal momentum fractions
are expressed in terms of rapidities and transverse momenta of $W$ bosons.

-
\begin{eqnarray}
x_1 &=& \sqrt{ {\bp_1^2 + m_W^{2} \over s}} e^{y_1} +  \sqrt{ {\bp_2^2 +
		m_W^2 \over s}} e^{y_2} 
\; , \nonumber \\
x_2 &=& \sqrt{ {\bp_1^2 + m_W^{2} \over s}} e^{-y_1} +  \sqrt{ {\bp_2^2 + m_W^2 \over s}} e^{-y_2} \, .
\end{eqnarray}

The elementary $\gamma \gamma \to W^+ W^-$ processes
in the Standard Model are shown in Fig.\ref{fig:diagrams}.

\begin{figure}
\begin{center}
\includegraphics[width=4cm,height=2.cm]{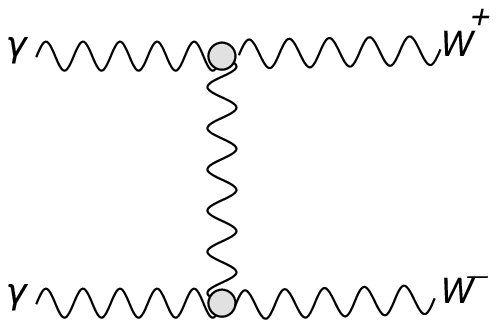}
\includegraphics[width=4cm,height=2.cm]{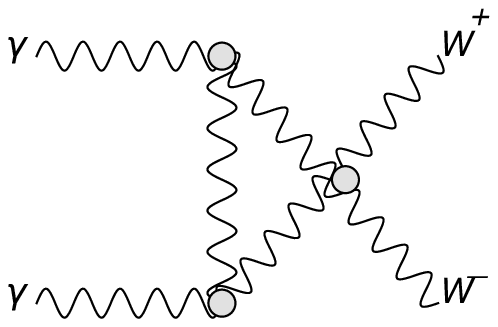}
\includegraphics[width=4cm,height=2.cm]{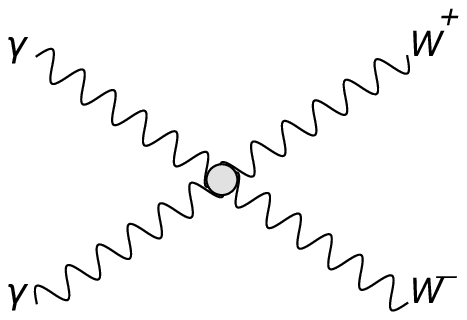}
\caption{
Different Feynman diagrams for photon-photon induced mechanisms for 
production of $W^+ W^-$ pairs.
}
\label{fig:diagrams}
\end{center}
\end{figure}


The elementary off-shell cross section
in (\ref{eq:kt-fact}) is written as:

\begin{eqnarray}
{d \sigma^*(p_1,p_2;\bq_1,\bq_2) \over dy_1 dy_2 d^2\bp_1 d^2\bp_2} 
= {1 \over 16 \pi^2 (x_1 x_2 s)^2} 
\sum_{\lambda_{W^+} \lambda_{W^-}} |M(\lambda_{W^+}, \lambda_{W^-})|^2 \, \delta^{(2)}(\bp_1 + \bp_2 - \bq_1 - \bq_2) \; .
\nonumber
\end{eqnarray}

Above the helicity-dependent off-shell matrix elements were calculated as:
\begin{eqnarray}
M(\lambda_{W^+} \lambda_{W^-} )&=& 
{ 1 \over |\qperp_1||\qperp_2|} 
\sum_{\lambda_1 \lambda_2} (\eperp(\lambda_1) \cdot \qperp_1) 
(\eperp^*(\lambda_2) \cdot \qperp_2) 
{\cal M}(\lambda_1, \lambda_2; \lambda_{W^+},\lambda_{W^-})  \nonumber \\
&=& { 1 \over |\qperp_1||\qperp_2|} \sum_{\lambda_1 \lambda_2} q_{\perp 1}^i q_{\perp 2}^j  \, e_i(\lambda_1) e_j^*(\lambda_2) 
{\cal M}(\lambda_1, \lambda_2; \lambda_{W^+},\lambda_{W^-})  \; .
\end{eqnarray}

Initial and final state helicity-dependent matrix elements were
discussed e.g. in \cite{Nachtmann:2005en}.
The $k_t$-factorization W-boson helicity dependent matrix elements
were calculated with the help of the above \cite{LSS2018}.

The unintegrated inelastic flux of photons is expressed as:
\begin{eqnarray}
{\cal{F}}^{\mathrm{in}}_{\gamma^* \leftarrow A} (z,\bq) &=& {\alpha_{\rm em} \over \pi} 
\Big\{(1-z) \Big( {\bq^2 \over \bq^2 + z (M_X^2 - m_p^2) + z^2 m_p^2  }\Big)^2  
{F_2(x_{\rm Bj},Q^2) \over Q^2 + M_X^2 - m_p^2}  \nonumber \\
&+& {z^2 \over 4 x^2_{\rm Bj}}  
{\bq^2 \over \bq^2 + z (M_X^2 - m_p^2) + z^2 m_p^2  }
{2 x_{\rm Bj} F_1(x_{\rm Bj},Q^2) \over Q^2 + M_X^2 - m_p^2} \Big\} \, ,
\end{eqnarray}
The main ingredients of the formula are $F_1$ and $F_2$ proton 
structure functions.

The unintegrated elastic flux of photons is expressed as:
\begin{eqnarray}
{\cal{F}}^{\mathrm{el}}_{\gamma^* \leftarrow A} (z,\bq) &=& {\aem \over \pi} 
\Big\{ (1-z)   \,\Big( {\bq^2 \over \bq^2 + z (M_X^2 - m_p^2) + z^2 m_p^2  }\Big)^2
{4 m_p^2 G_E^2(Q^2) + Q^2  G_M^2(Q^2) \over 4m_p^2 + Q^2} \nonumber \\
&+& {z^2 \over 4}  {\bq^2 \over \bq^2 + z (M_X^2 - m_p^2) + z^2 m_p^2  } G_M^2(Q^2) \Big\} \; .
\nonumber \\
\end{eqnarray}
In this case the main ingredients are $G_E$ and $G_M$ electromagnetic
form factors of proton.

To calculate inelastic fluxes of photons one needs numerical
representation of structure functions of protons.
Different parametrizations of $F_2$ structure functions are available
in the literature, see e.g. \cite{Abramowicz:1997ms,SU,MNSZ2017}.

\section{Results}

The integrated cross sections obtained in our approach are collected 
in Table 1.

\begin{table}[tbp]
\centering
\begin{tabular}{|c|c|c|}
\hline
contribution               &  8 TeV  & 13 TeV  \\
\hline
        LUX-like             &      &          \\

$\gamma_{el} \gamma_{in}$  & 0.214 & 0.409   \\
$\gamma_{in} \gamma_{el}$  & 0.214 & 0.409   \\
$\gamma_{in} \gamma_{in}$  & 0.478 & 1.090   \\
\hline
      ALLM97 F2      &      &          \\

$\gamma_{el} \gamma_{in}$  & 0.197 & 0.318   \\
$\gamma_{in} \gamma_{el}$  & 0.197 & 0.318    \\
$\gamma_{in} \gamma_{in}$  & 0.289 & 0.701    \\
\hline
      SU F2 &      &          \\

$\gamma_{el} \gamma_{in}$  & 0.192 & 0.420    \\
$\gamma_{in} \gamma_{el}$  & 0.192 & 0.420    \\
$\gamma_{in} \gamma_{in}$  & 0.396 & 0.927    \\
\hline
    LUXqed collinear &      &          \\

$\gamma_{in+el}$ $ \gamma_{in+el}$  & 0.366 & 0.778    \\
\hline
    MRST04 QED collinear &      &          \\

$\gamma_{el} \gamma_{in}$  & 0.171 & 0.341    \\
$\gamma_{in} \gamma_{el}$  & 0.171 & 0.341    \\
$\gamma_{in} \gamma_{in}$  & 0.548 & 0.980    \\
\hline
    Elastic- Elastic&      &          \\
$\gamma_{el} \gamma_{el}$ (Budnev)  & 0.130 & 0.273   \\

$\gamma_{el} \gamma_{el}$  (DZ)  & 0.124 & 0.267  \\

\hline

\end{tabular}
\caption{Cross sections (in $p b$) for different contributions 
and different $F_2$ structure functions: LUX, ALLM97 and SU, 
compared to the relevant collinear distributions with MRST04 QED 
and LUXqed distributions. 
}
\label{table:1}
\end{table}

Without any gap survival effects:
\begin{equation}
\sigma(inel.-inel.) > \sigma(inel.-el.)+\sigma(el.-inel.) > \sigma(el.-el.) \; .
\label{inclusive_cs_ordering}
\end{equation}

\begin{figure}
\begin{center}
\includegraphics[height=5.0cm]{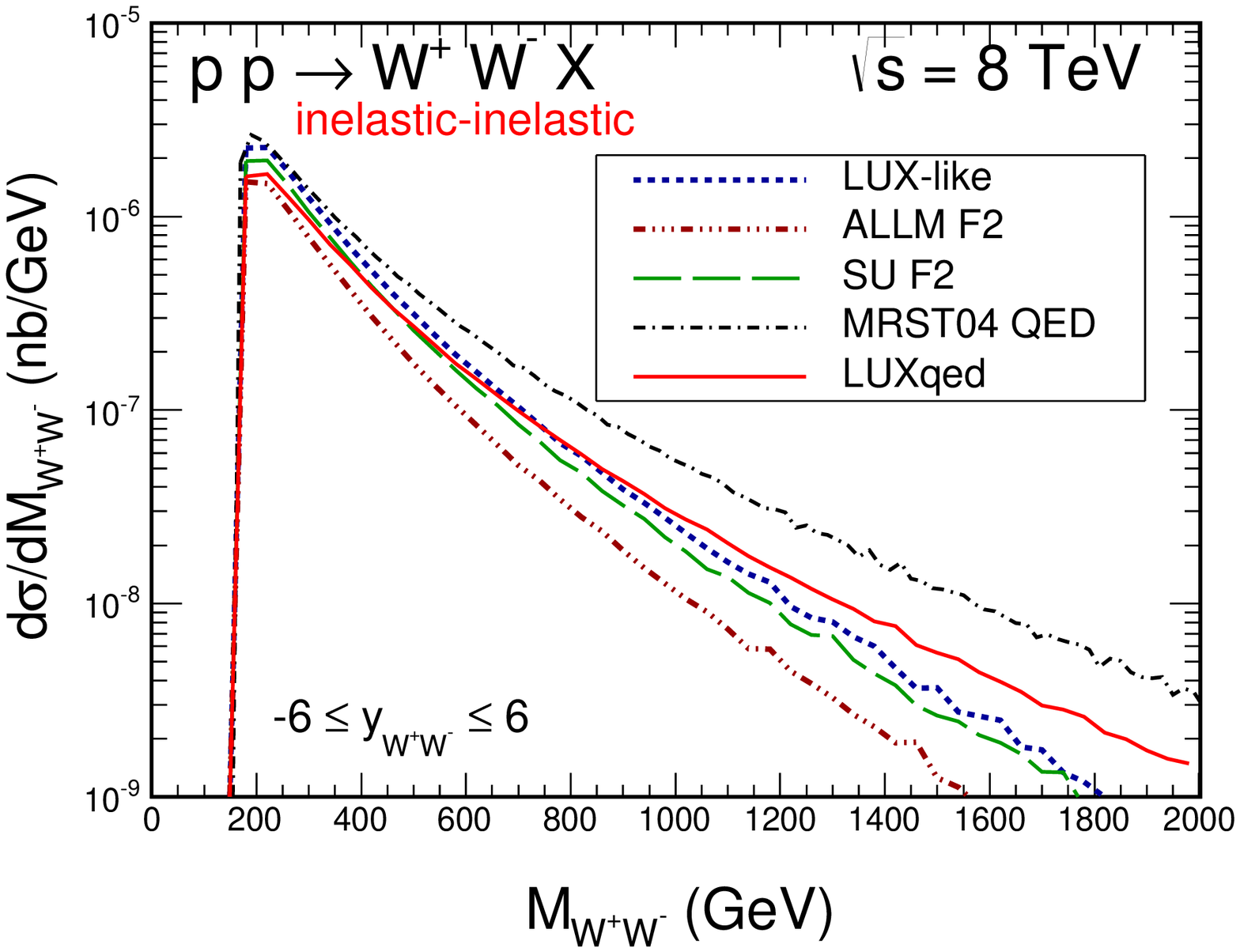}
\includegraphics[height=5.0cm]{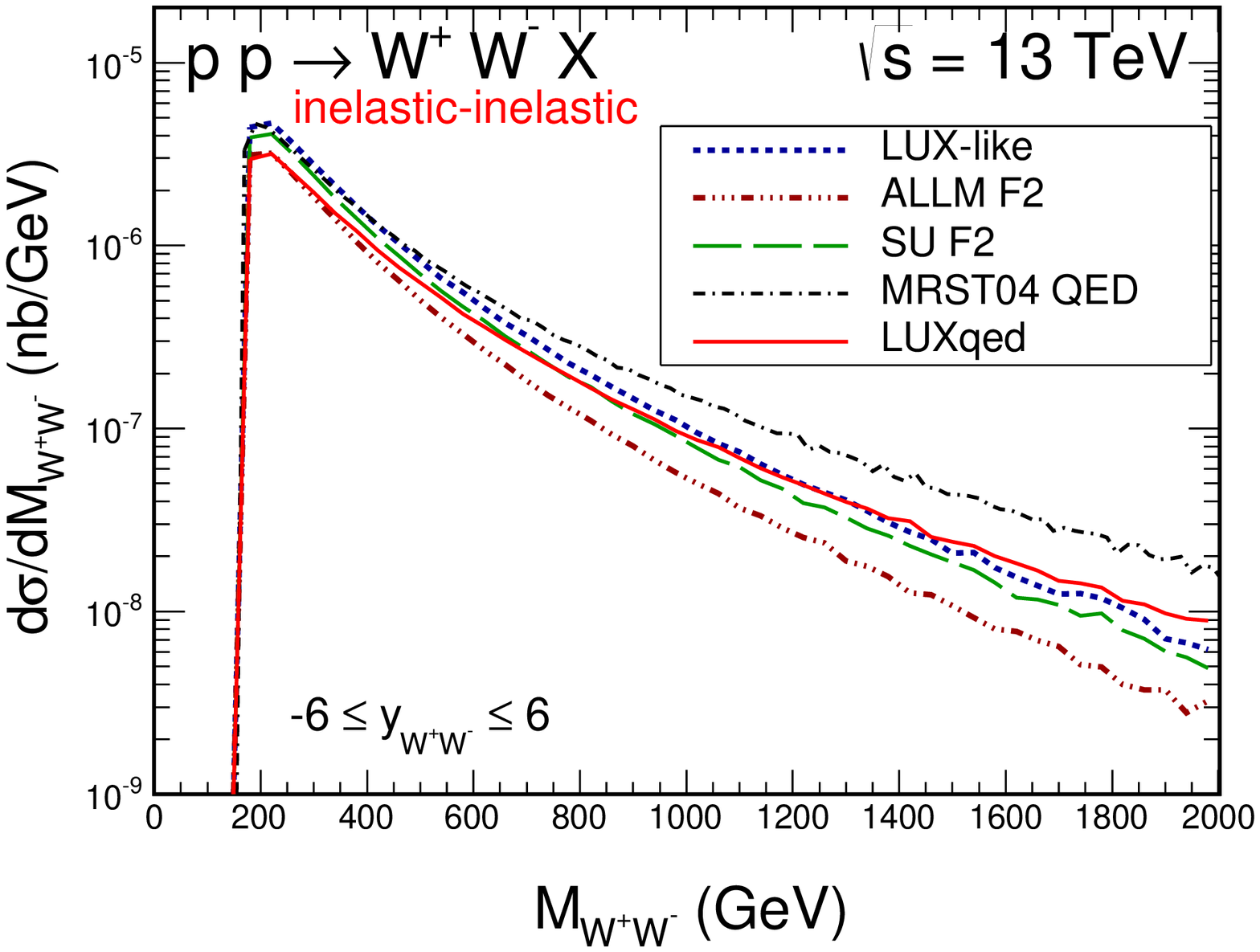}
\caption{$M_{WW}$ invariant mass distribution for double dissociative 
contribution obtained with different parametrizations of structure functions.}
\end{center}
\label{fig:dsig_dM}
\end{figure}

Many differential distributions were calculated in \cite{LSS2018}.
Here, in Fig.\ref{fig:dsig_dM}, we show only invariant mass distribution 
for double dissociation processes (inelastic-inelastic) for different 
parametrizations of the structure functions from the literature.

The $k_t$-factorization result is similar to the collinear one
for the same structure function (LUX-like).
The rather old MRST04-QED collinear approach \cite{Martin:2004dh}
predicted larger cross section. The reasons were discussed in \cite{LSS2018}.

\begin{figure}
\begin{center}
\includegraphics[height=6.0cm]{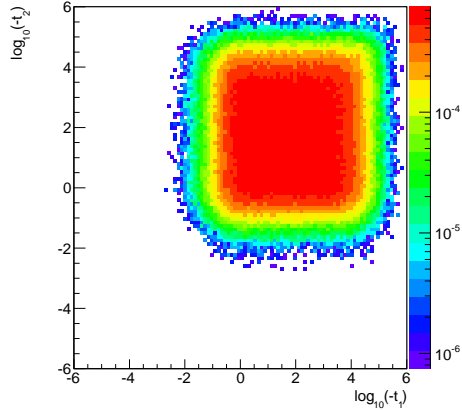}
\caption{Two-dimensional distribution in
  $(log_{10}(Q_1^2),log_{10}(Q_2^2))$
for double dissociative process.}
\end{center}
\label{fig:virtualities}
\end{figure}

As an example in Fig.\ref{fig:virtualities} we show distribution in
virtualities of photons. Rather large virtualities of photons come into game.
The large virtualities of photons seem to contradict collinear approach.

Our formalism allows to calculate contributions depending on
helicities of $W^+$ and $W^-$ bosons. The results are collected in
Table 2 for two different collision energies.
Clearly the $TT$ contribution dominates.

\begin{table}[tbp]
\centering
\begin{tabular}{|c|c|c|c|}
\hline
contribution               &   8 TeV   &  13 TeV        \\
\hline        
TT                         &   0.405   &  0.950         \\
\hline
LL                         &   0.017   &  0.046         \\
\hline
LT + TL                & 0.028 + 0.028 & 0.052 + 0.052  \\            
\hline
SUM                        & 0.478 &  1.090   \\  
\hline
 
\end{tabular}
\caption{Contributions of different polarizations of $W$ bosons
for the inelastic-inelastic component for the LUX-like structure function. 
The cross sections are given in $p b$. 
}
\end{table}

\begin{figure}
  \centering
  \includegraphics[width=.40\textwidth]{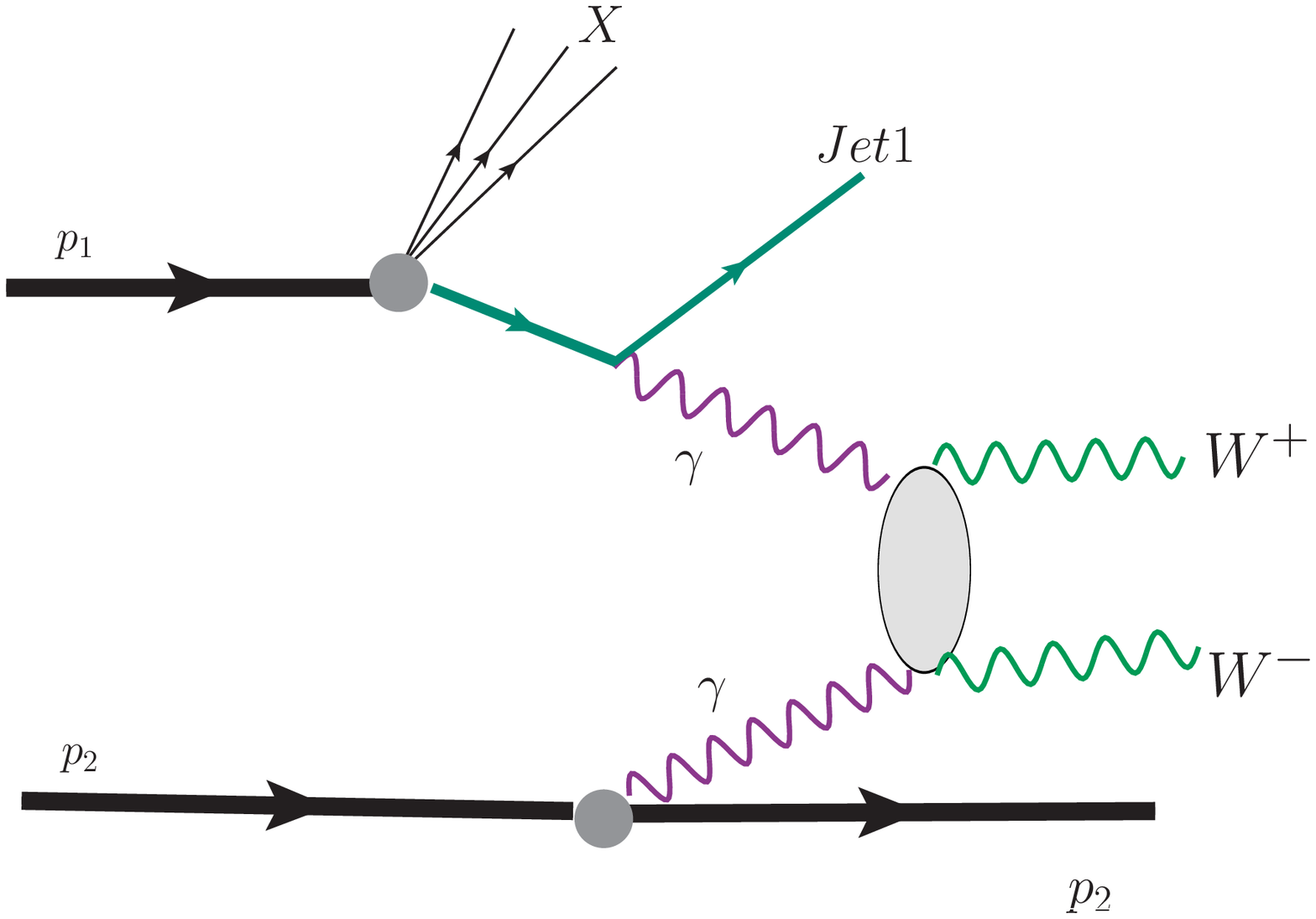}
  \includegraphics[width=.40\textwidth]{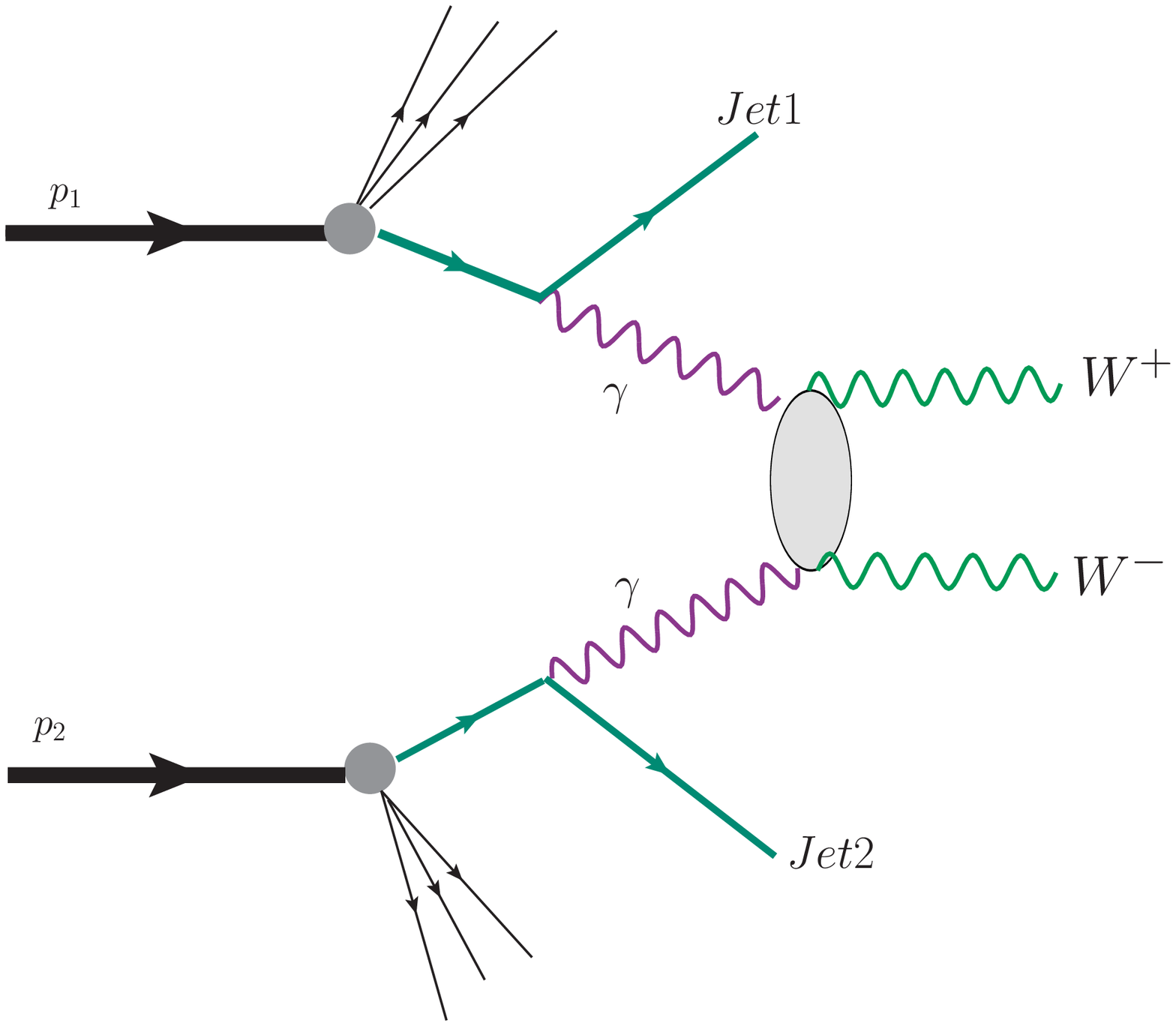}
  \caption{Schematic representation of the single and double
    dissociative mechanisms. Jets are shown explicitly.
 }
  \label{fig:diagrams_dissociation}
\end{figure}

The remnant fragmentation \cite{LFSS2019a} was done with the help of
PYTHIA 8 program.
Including only parton (jet) emission is already a quite good approximation.

The gap survival probability for single dissociative process is
calculated as:
\begin{equation}
S_R(\eta_{\rm cut}) = 1 - {1 \over \sigma}
\int_{-\eta_{\rm cut}}^{\eta_{\rm cut}} \frac{{\rm d}\sigma}{{\rm d}
    \eta_{\rm jet}} {\rm d} \eta_{\rm jet} \; .
\label{parton_model_gap_survival}
\end{equation}

A schematic representation of remnant fragmentation(s) with explicit jet
is shown in Fig.\ref{fig:diagrams_dissociation}.
Jet emissions were considered also in \cite{Durham}.

The gap survival factor associated with jet emission is shown in 
Fig.6. 

\begin{figure}
\centering
\includegraphics[width=.63\textwidth]{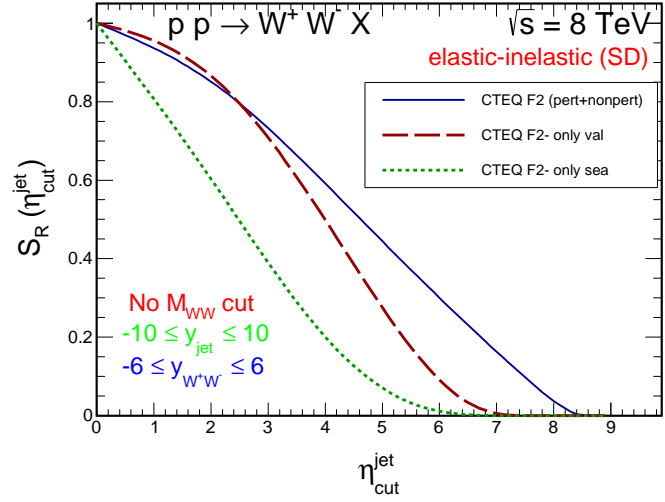}
\caption{Gap survival factor for single dissociative process
associated with the jet emission. 
The solid line is for the full model, the dashed line for 
the valence contribution and the dotted line for the sea contribution.
}
 \label{fig:S_R_partonic}
\end{figure}

\begin{figure}
\begin{center}
\includegraphics[width=.50\textwidth]{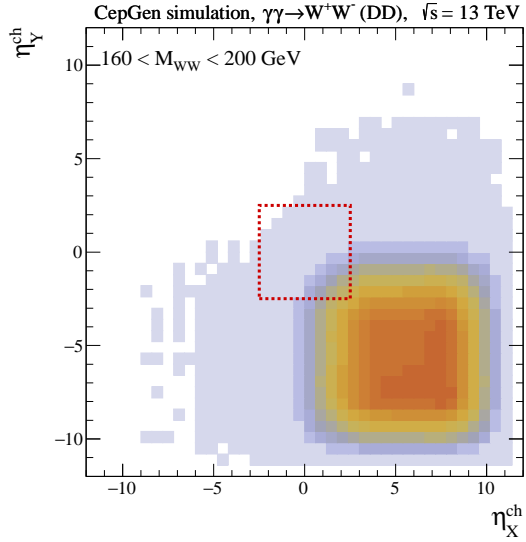}
\caption{Two-dimensional ($\eta^{\rm ch}_{X},\eta^{\rm ch}_{Y}$) distribution
for a selected window of $M_{WW}$.
The square shows pseudorapidity coverage of ATLAS or CMS inner tracker.}
\end{center}
\label{fig:dsig_deta1deta2_MWW_windows}
\end{figure}

{\tiny
\begin{table}[tbp]
\centering
\begin{tabular}{|c|c|c|c|c|c|c|}
\hline
                       & $8~TeV$  & $13~TeV$ & $8~TeV$  & $13~TeV$  & $8~TeV$ & $13~TeV$\\
\hline
$(2 M_{WW}, 200~GeV)$  & 0.763(2)  & 0.769(2)  & 0.582(4)  & 0.591(4)   & 0.586(1) & 0.601(2)\\
\hline
$(200, 500~GeV)$       & 0.787(1)  & 0.799(1)  & 0.619(2)  & 0.638(2)   & 0.629(1) & 0.649(1)\\
\hline
$(500, 1000~GeV)$      & 0.812(2)  & 0.831(2)  & 0.659(3)  & 0.691(3)   & 0.673(2) & 0.705(2)\\
\hline
$(1000, 2000~GeV)$     & 0.838(7)  & 0.873(5)  & 0.702(12) & 0.762(8)   & 0.697(5) & 0.763(6)\\
\hline
full range             & 0.782(1)  & 0.799(1)  & 0.611(2)  & 0.638(2)   & 0.617(1) & 0.646(1)\\
\hline
\end{tabular}
\caption{Average rapidity gap survival factors: 
$S_{R,SD}(|\eta^{ch}|<2.5)$,
$\left(S_{R,SD}\right)^2(|\eta^{ch}|<2.5)$,
$S_{R,DD}(|\eta^{ch}|<2.5)$
related to remnant fragmentation
for {\it single dissociative} and {\it double dissociative} contributions
for different ranges of $M_{WW}$.
}
\label{tab:gsf_sd}
\end{table}
}
Fig.7 illustrates how gap survival 
factor is destroyed by particle (hadron) emission for 
double dissociative process.

We find (see also Table 1)
\begin{equation}
S_{R,DD} \approx \left(S_{R,SD}\right)^2  \; .
\label{factorisation}
\end{equation}
Such an effect is naively expected when 
the two fragmentations are independent, which is the case
by the model construction.
The soft processes will most probably violate the factorisation.
There is, however, no formalism which allows to calculate the gap
survival probabability for these processes as a function of rapidity
gap window.
So far we have not included the soft gap survival factors.
They are relatively easy to calculate only for double elastic (DE)
contribution \cite{LS_HH}.
For the ``soft'' gap survival factors we expect:
\begin{equation}
S_{soft}(DD) < S_{soft}(SD) < S_{soft}(DE) \; .
\end{equation}

Finally we wish to show also similar results for $p p \to t {\bar t}$
reaction. In Table 3 we show integrated cross sections for different
categories of processes. Rather small cross sections are obtained.
It is not clear at present whether such a process can be identified
experimentally.

\begin{table}[tbp]
\centering
\begin{tabular}{|c|c|c|}
\hline
Contribution          &  No cuts & $y_{\rm jet}$ cut\\
\hline
elastic-elastic       &  0.292 &  0.292 \\
elastic-inelastic     &  0.544 &  0.439 \\
inelastic-elastic     &  0.544 &  0.439 \\
inelastic-inelastic   &  0.983 &  0.622 \\
\hline
all contributions     &  2.36  &  1.79 \\
\hline
\end{tabular}
\caption{Cross section for $t {\bar t}$ production in fb at $\sqrt{s}$ = 13 TeV for 
different components (left column) and the same when the extra condition
on the outgoing jet 
$ |y_{\rm jet}| > 2.5$
is imposed.}
\label{tab:sig_tot}
\end{table}

As an example we show $t {\bar t}$ invariant mass distribution
for inclusive case as well as when extra veto on (mini)jet
is imposed. The inclusion of rapidity gap veto reduces the cross
section. Whether the cross section corresponding to the photon-photon
fusion can be measured requires special dedicated studies.

\begin{figure}
  \centering
  \includegraphics[width=.48\textwidth]{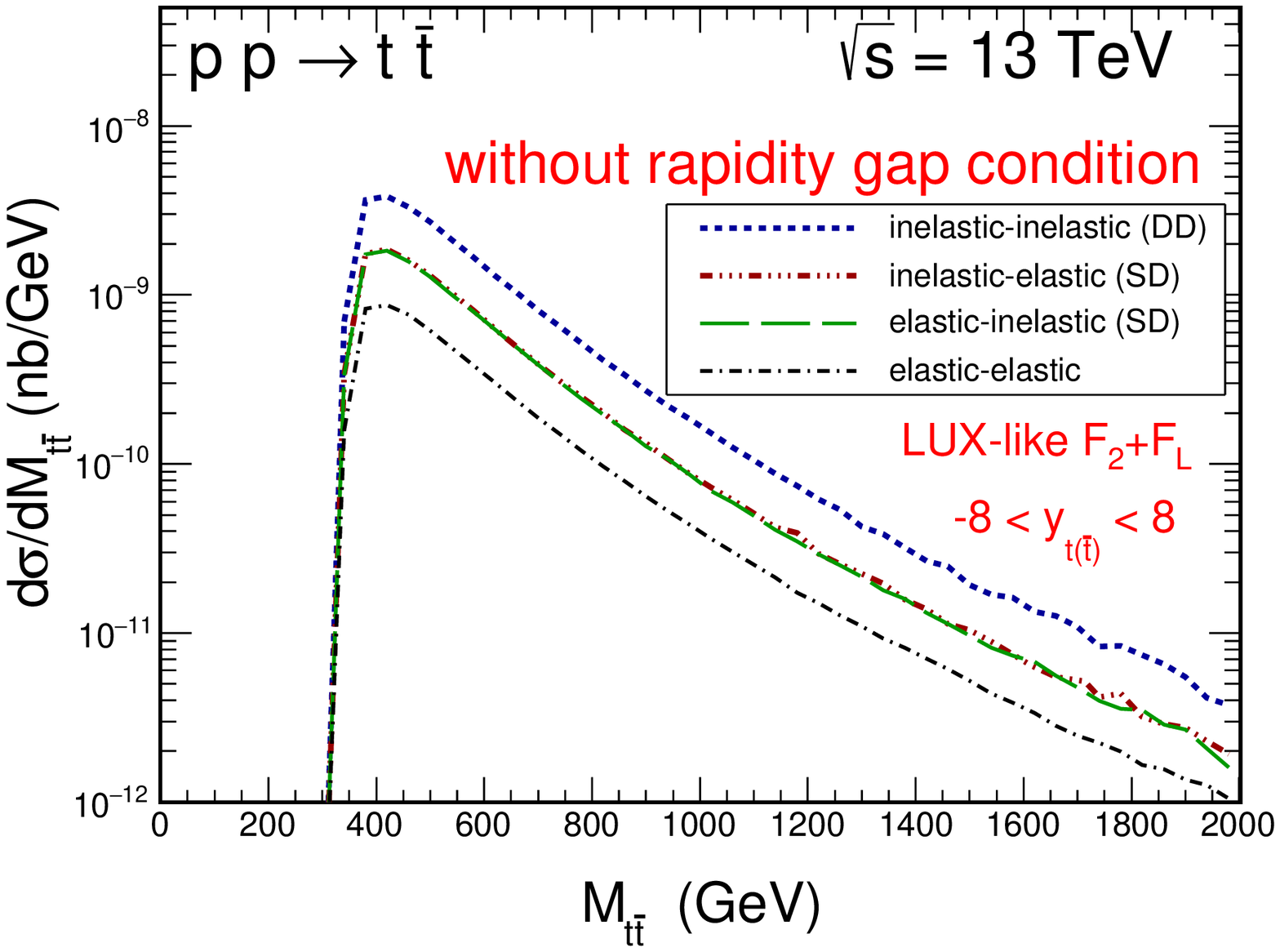}
  \includegraphics[width=.48\textwidth]{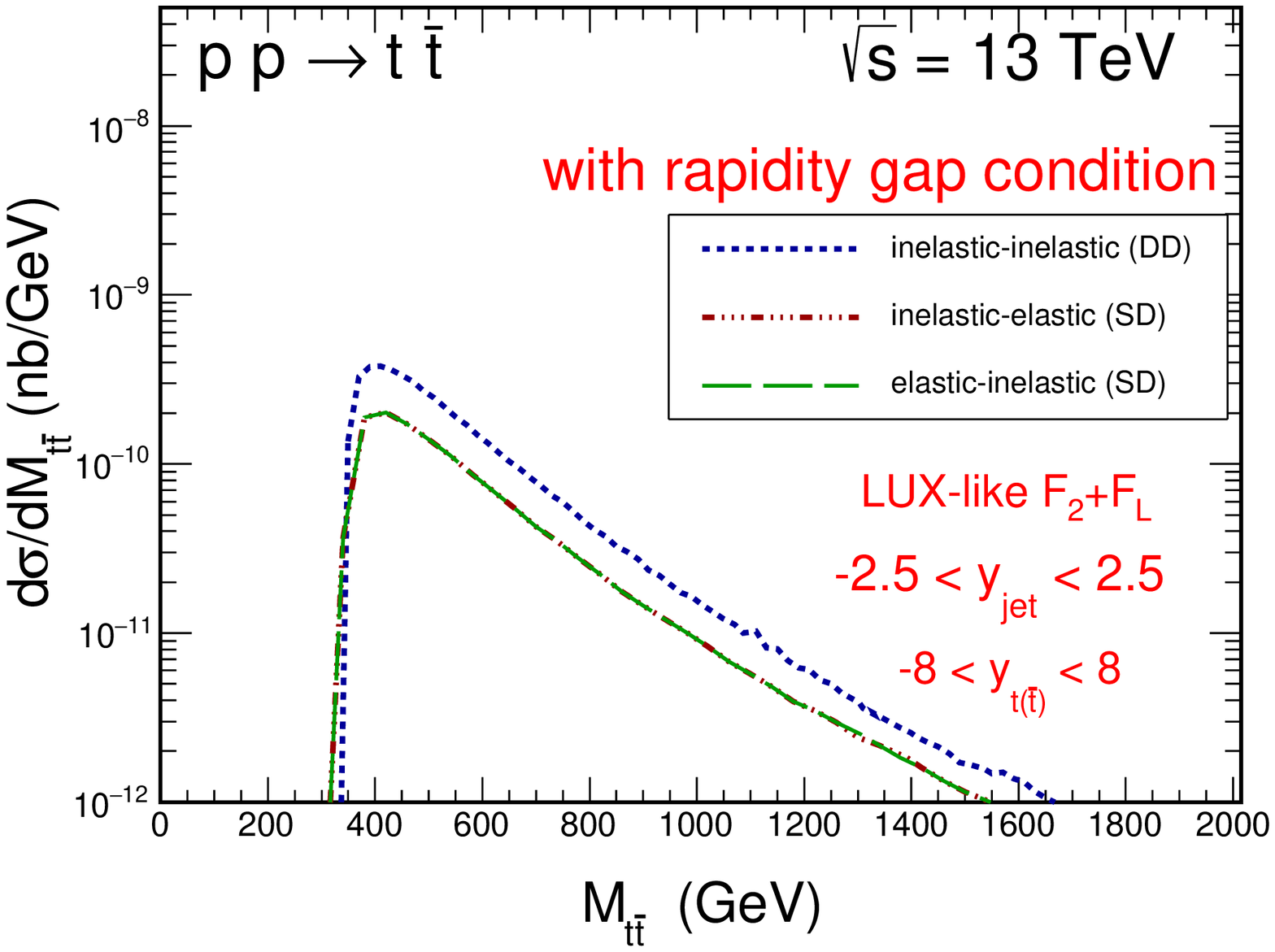}
  \caption{$t \bar t$ invariant mass distribution for different 
components defined in the figure. The left panel is without imposing
the condition on the struck quark/antiquark and the right panel includes
the condition.
}
\label{fig:dsig_dMttbar}
\end{figure}

\section{Conclusions}

Helicity-dependent matrix elements for $\gamma^* \gamma^* \to W^+ W^-$ 
(off-shell photons) have been derived and
used in the calculation of cross sections for $p p \to W^+ W^-$ reaction.
We have obtained cross section of about 1 pb for the LHC energies.
This is about 2 \% of the total integrated cross section dominated 
by the quark-antiquark annihilation and gluon-gluon fusion.
Different combinations of the final states 
(elastic-elastic, elastic-inelastic, inelastic-elastic, inelastic-inelastic)
have been considered.
The unintegrated photon fluxes were calculated based on modern 
parametrizations of the proton structure functions from the literature.
Several differential distributions in $W$ boson transverse momentum
and rapidity, $WW$ invariant mass, transverse momentum of 
the $WW$ pair, mass of the remnant system have been presented.
Several correlation observables have been studied. 
Large contributions from the regions of 
large photon virtualities $Q_1^2$ and/or $Q_2^2$ 
have been found putting in question the reliability 
of leading-order collinear-factorization approach. 
We have presented a decomposition of the cross section into
different polarizations of both $W$ bosons.
It has been shown that the TT (both $W$ transversly polarized)
contribution dominates and constitutes more than 80 \% of the total cross section.
The $LL$ (both $W$ longitudinally polarized) contribution 
is interesting in the context of studying
$W W$ interactions or searches beyond the Standard Model.
We have quantifield the effect of inclusion of 
longitiudinal structure function
into the transverse momentum dependent fluxes of photons. 
A rather small, approximataly $M_{WW}$~-~independent, effect was found.

We have discussed the quantity called ``remnant gap survival factor'' for 
the $pp \to W^+ W^-$ reaction initiated via photon-photon fusion.
We have calculated the gap survival factor
for single dissociative process on the parton level.
In such an approach the outgoing parton (jet/mini-jet) is responsible
for destroying the rapidity gap.
We have found that the hadronisation only mildly modifies
the gap survival factor calculated on the parton level.
This may justify approximate treatment of hadronisation of remnants.
We have found different values for double and single
dissociative processes.
In general, $S_{R,DD} < S_{R,SD}$ and 
$S_{R,DD} \approx (S_{R,SD})^2$.
We expect that the factorisation observed here for
the remnant dissociation and hadronisation will be violated
when the soft processes are explicitly included.
The larger $\eta_{\rm cut}$ (upper limit on charged
particles pseudorapidity), the smaller rapidity gap survival factor
$S_R$. This holds both for the double and the single dissociation.
The present approach is a first step towards 
a realistic modelling of gap survival in photon induced interactions 
and definitely requires further detailed studies and comparisons 
to the existing and future experimental data.

We have also calculated cross sections for $t \bar t$ production
via $\gamma \gamma$ mechanism in $pp$ collisions
including photon transverse momenta
and using modern parametrizations of proton structure functions.
The contribution to the inclusive $t \bar t$ is only about 2.5 fb.
We have found
$\sigma_{tt}^{ela-ela} < \sigma_{tt}^{SD} < \sigma_{tt}^{DD}$.
We have calculated several differential distributions. 
Some of them are not accessible in standard equivalent photon approximation.
As for $W^+ W^-$ production we have shown that rather large photon 
virtualities come into the game.

\end{document}